%% file: icassp26_javid_cameraready.tex
\documentclass{article}
\usepackage{spconf}
\usepackage{microtype}      % microtypography
\usepackage[utf8]{inputenc} % allow utf-8 input
\usepackage[T1]{fontenc}    % use 8-bit T1 fonts
\usepackage{xurl}
\usepackage{hyperref}       % hyperlinks
\usepackage{booktabs}       % professional-quality tables
\usepackage{subcaption}    % for subfigure environment
\usepackage{graphicx}      % for \includegraphics
\usepackage{amsfonts}       % blackboard math symbols
\usepackage{nicefrac}       % compact symbols for 1/2, etc.
\usepackage{xcolor}         % colors
\usepackage{amsmath}
\usepackage{amssymb}
\usepackage{algorithm}
\usepackage{algpseudocode}
\usepackage{breakurl}
\usepackage[english]{babel}
\usepackage{amsthm}
\usepackage{tabularx,booktabs,siunitx}
\usepackage{adjustbox}  % in your preamble
\usepackage[english]{babel}
\usepackage{amsthm}

\newtheorem*{remark}{Remark}

\input{macros.tex}
\input{acronyms.tex}

\def\BibTeX{{\rm B\kern-.05em{\sc i\kern-.025em b}\kern-.08em
		T\kern-.1667em\lower.7ex\hbox{E}\kern-.125emX}}

\definecolor{cadmiumgreen}{HTML}{40916c}

\begin{document}
	
	\title{Environment-Aware MIMO Channel Estimation in Pilot-Constrained Upper Mid-Band Systems}
	\name{Seyed Alireza Javid and  Nuria Gonz\'alez-Prelcic}
	\address{University of California San Diego, USA}

	\maketitle
	
	\begin{abstract}
		Accurate \ac{MIMO} channel estimation is critical for next-generation wireless systems, enabling enhanced communication and sensing performance. Traditional model-based channel estimation methods suffer, however, from performance degradation in complex environments with a limited number of pilots, while purely data-driven approaches lack physical interpretability, require extensive data collection, and are usually site-specific. This paper presents a novel \ac{PINN} framework that combines model-based channel estimation with a deep network to exploit prior information about the propagation environment and achieve superior performance under pilot-constrained scenarios. The proposed approach employs an enhanced U-Net architecture with cross-attention mechanisms to fuse initial channel estimates with \ac{RSS} maps to provide refined channel estimates. Comprehensive evaluation using realistic ray-tracing data from urban environments demonstrates significant performance improvements, achieving over 5 dB gain in \ac{NMSE} compared to state-of-the-art methods, with particularly strong performance in pilot-limited scenarios and robustness across different frequencies and environments with only minimal fine-tuning. The proposed framework maintains practical computational complexity, making it viable for massive \ac{MIMO} systems in upper mid-band frequencies.
	\end{abstract}
	
	\begin{keywords}
		Physics-informed neural network, channel estimation, upper mid-band, MIMO.\end{keywords}
	
	\section{Introduction}
	\vspace*{-3mm}
	Channel state information (CSI) is exploited by several functional blocks in a \ac{MIMO} communication system, including precoding and combining, link adaptation, or equalization \cite{4267831}. High accuracy CSI is especially important when large arrays are exploited, multiple users (MU) are served, the channel is wideband, or the communication signal is also exploited for localization or sensing \cite{palacios2023separable}. In these scenarios, CSI inaccuracies may lead to beam misalignment, multiuser interference, poor equalization, or reduced sensing accuracy.
	
	Traditional model-based channel estimation techniques for large-array \ac{MIMO} systems--as those expected in the upper mid-band--often rely on pilots. However, as the number of antennas increases, they require excessive training and incur high computational complexity, which leads to performance degradation under overhead constraints or in complex propagation environments \cite{palacios2022multidimensional}.
	%Traditional model-based channel estimation techniques for large-array MIMO systems often rely on pilots, but as the number of antennas increases, they require excessive training and incur high computational complexity, leading to performance degradation under overhead constraints or in complex propagation environments \cite{palacios2022multidimensional, palacios2023separable, venugopal2017channel, rodriguez2018frequency}.
	Prior work has also proposed black-box, fully data-driven approaches \cite{sattari2024full, helmy2025low, zhou2025generative} that often suffer from limited physical interpretability, demand large-scale data collection, or tend to generalize poorly beyond the specific site where they were trained.
	Another line of work combines model-based priors with data-driven learning \cite{jin2025near}. However, these combined approaches often utilize the black-box structure of neural networks along with simplistic channel models and unrealistic statistical data for training. Consequently, performance often degrades under domain (site) shift, and substantial site-specific data are required for training when the deployment location changes. Finally, the exploitation of diffusion models in the recent channel estimation literature \cite{jin2025near, zhou2025generative}  will also introduce a high overhead in the backward process for generating the channels, and cannot be used in real-time communication systems. 
	
	This paper presents the first \ac{PINN} framework for \ac{MIMO} channel estimation in the upper mid-band under pilot-constrained scenarios. Unlike prior work that uses PINNs for path 
	loss \cite{jiang2024physics}, we integrate \ac{RSS} maps with LS-based 
	channel estimates through a physics-informed U-Net architecture enhanced with transformers and cross-attention to integrate environmental propagation information with coarse channel information. We validate the design using realistic ray-tracing datasets across frequencies and environments, demonstrating strong generalization. We make the data set available to the research community.   
	
	\vspace*{-3mm}
	\section{Physical and communication models} \label{model}
	\vspace*{-2mm}
	\subsection{Physical model} 
	\vspace*{-2mm}Maxwell’s equations govern wireless propagation by linking electric and magnetic fields with material properties \cite{ida2015engineering}. %Under time-harmonic assumptions, they reduce to an inhomogeneous wave equation whose solution via Green’s functions expresses the field as a superposition of contributions from sources and medium inhomogeneities, determined by parameters such as permittivity, permeability, and conductivity. 
	After computing the electric field distribution using numerical methods, the total received power at a given location, or \ac{RSS},  can be determined from the electric field. Specifically, the received power is proportional to the squared magnitude of the electric field integrated over the receiving antenna's effective area. Mathematically, the RSS accounting for all \ac{MPC}, can be expressed as \cite{fu2023fast}
	\vspace*{-3mm}
	\begin{equation}
		\text{RSS}(\mathbf{r}) = \frac{\lambda^2}{8 \pi \eta_0} \left| \sum_{\ell=1}^{P} E_{\ell}(\mathbf{r}) \right|^2,
		\label{eq:received_power}
	\end{equation}
	where $P$ denotes the number of propagation paths, $\lambda$ is the wavelength, $\eta_0$ is the intrinsic impedance of free space, and $E_{\ell}$ represents the complex amplitude of the electric field associated with the $\ell$-th path at the receiver location $\br$. Using $E(\mathbf{r}) = \sum_{\ell=1}^{P} E_{\ell}(\mathbf{r})$, this relates to the total electric field. This expression captures the coherent superposition of all paths, including their respective amplitudes and phases.
	%This relationship forms the theoretical basis for connecting field simulations to practical received signal strength measurements in wireless communication systems.
	
	\vspace*{-2mm}
	\subsection{Communication channel model} 
	\vspace*{-1mm}We consider a communication system operating in the upper mid-band (7-24 GHz) in an urban environment. The \ac{BS} equipped with a \ac{URA} of size $N_t = N_t^x \times N_t^y$ is at the top of a building, and the user equipment (UE) on the ground operating at the same frequency is equipped with a \ac{URA} consisting of $N_r = N_r^x \times N_r^y$ elements. The frequency selective channel for the $d$-th  channel tap  consists of ${P}$ paths, and   can be modeled as
	\vspace*{-2mm}
	\begin{equation}
		\label{channel}
		\mathbf{H}_d=\sum_{\ell=1}^{P}\alpha_{\ell} f_{\mathrm{p}}\left(d T_{\mathrm{s}}-\left(t_{\ell}-t_{\mathrm{off}}\right)\right)\mathbf{a}_{\mathrm{r}}\left(\theta_{\ell}^{\mathrm{az}}, \theta_{\ell}^{\mathrm{el}}\right) \mathbf{a}_{\mathrm{t}}\left(\phi_{\ell}^{\mathrm{az}}, \phi_{\ell}^{\mathrm{el}}\right),
	\end{equation}
	where  $T_\textrm{s}$ is the sampling interval, $t_{\mathrm{off}}$ is the clock offset between the transmitter and receiver, $f_{\mathrm{p}}(.)$ is the filtering function that factors in filtering effects in the system, $\alpha_{\ell}$ and $t_{\ell}$ are the complex gain and the \ac{ToA} of the $l$-th path, $\mathbf{a}_{\mathrm{r}}\left(\theta_{\ell}^{\mathrm{az}}, \theta_{\ell}^{\mathrm{el}}\right)$ represents the receiver array response evaluated at the azimuth and elevation \ac{AoA}, denoted as $\theta_{\ell}^{\mathrm{az}}$ and $ \theta_{\ell}^{\mathrm{el}}$ respectively, and $ \mathbf{a}_{\mathrm{t}}\left(\phi_{\ell}^{\mathrm{az}}, \phi_{\ell}^{\mathrm{el}}\right)$ is the transmitter array response evaluated at the azimuth and elevation \ac{AoD}, denoted as $\phi_{\ell}^{\mathrm{az}}$ and $ \phi_{\ell}^{\mathrm{el}}$ respectively. The array responses can be written as
	\vspace*{-3mm}
	\begin{subequations}
		\begin{align}
			\mathbf{a}_{\mathrm{r}}\left(\theta^{\mathrm{az}}, \theta^{\mathrm{el}}\right)=\mathbf{a}\left(\theta^{\prime \prime}, \theta^{\perp}\right)=\mathbf{a}\left(\theta^{\prime \prime}\right) \otimes \mathbf{a}\left(\theta^{\perp}\right) \\
			\mathbf{a}_{\mathrm{t}}\left(\phi^{\mathrm{az}}, \phi^{\mathrm{el}}\right)=\mathbf{a}\left(\phi^{\prime \prime}, \phi^{\perp}\right)=\mathbf{a}\left(\phi^{\prime \prime}\right) \otimes \mathbf{a}\left(\phi^{\perp}\right)
		\end{align}
	\end{subequations}
	where $\theta^{\prime \prime}=\cos \left(\theta^{\mathrm{el}}\right) \sin \left(\theta^{\mathrm{az}}\right), \theta^{\perp}=\sin \left(\theta^{\mathrm{el}}\right), \phi^{\prime \prime}=$ $\cos \left(\phi^{\mathrm{el}}\right) \sin \left(\phi^{\mathrm{az}}\right), \phi^{\perp}=\sin \left(\phi^{\mathrm{el}}\right)$, and $\mathbf{a}(\cdot)$ is the steering vector where $[\mathbf{a}(\vartheta)]_n=e^{-j \pi(n-1) \vartheta}$ assuming a halfwavelength element spacing.
	
	% THE CONNECTION
	\vspace*{-2mm}
	\subsection{Connecting the physical and the channel models} The electromagnetic analysis models propagation by explicitly computing the electric field from environmental properties. In contrast, communication models approximate the channel as a finite set of paths, each with delay, angle, and complex gain, represented in the \ac{MIMO} channel matrix $\{\mathbf{H}_d\}_{d=1}^D$ over $D$ delay taps.
	%Each matrix $\mathbf{H}_d \in \mathbb{C}^{N_r \times N_t}$ encapsulates the collective effect of all multipath components arriving at delay $d$, as sampled at the antenna elements:
	%\begin{equation}
	%\mathbf{y}_d = \mathbf{H}_d \mathbf{x}_d + \mathbf{n}_d,
	%\end{equation}
	%where $\mathbf{x}_d$ is the transmit vector and $\mathbf{y}_d$ the received vector at delay tap $d$. 
	The physically computed field superposition at each receiver location thus defines the per-tap response in the wideband \ac{MIMO} channel. Given the transmit power $P_T$, the RSS can be calculated per tap, or integrated across all taps, using the \ac{MIMO} channel matrices, as
	\vspace*{-3mm}
	\begin{equation}
		\text{RSS} =  P_T\sum_{d=1}^{D} \mathbb{E}\left[\|\mathbf{H}_d\|_F^2\right],
	\end{equation}
	or, equivalently, from the summed field strengths at each tap. 
	
	\vspace*{-3mm}
	\section{RSS-aware channel estimation} \label{nndisc}
	\vspace*{-2mm}
	This section describes the proposed  \ac{PINN}  framework for low overhead channel estimation. It exploits a few pilots to obtain a coarse channel estimate which is later fused  with RSS information to create a refined channel estimate. 
	\vspace*{-3mm}
	\subsection{Coarse channel estimation} 
	\vspace*{-2mm}The first step involves evaluating a coarse channel estimation. To simplify our approach and avoid unnecessary complications, we utilize a basic variation of \ac{LS} estimation \cite{ozdemirchannel}. However, the proposed \ac{PINN} framework can operate with other initial channel estimation methods. %A detailed formulation of this stage can be found in Appendix~\ref{ines}.
	
	The frequency-selective \ac{MIMO} channel including all taps is modeled as a 3-dimensional complex tensor $\mathbf{H} \in \mathbb{C}^{D \times N_{\text{r}} \times N_{\text{t}}}$ created by the concatenation of $D$ channel matrices as in \eqref{channel}. The received signal can be written as
	\vspace*{-3mm}
	\begin{equation}
		\mathbf{r}[n]= \sum_{d=0}^{D-1} \mathbf{H}_d \mathbf{s}[n-d]+\mathbf{v}[n],
	\end{equation}
	where $\mathbf{s}[n] \in \mathbb{C}^{N_{\text{t}}\times 1}$ is the transmitted pilot vector and $\mathbf{n}_{d} \sim \mathcal{CN}(0, \sigma_n^2 \mathbf{I})$ is the additive white Gaussian noise. We adopt a simple \ac{LS} estimator in the frequency domain using OFDM pilots. After LS estimation at pilot subcarriers, magnitude and phase are linearly interpolated to recover the full channel, followed by an inverse FFT to obtain the time-domain taps. This method provides a practical baseline with good initial accuracy for refinement by our physics-informed network.
	
	\vspace*{-3mm}
	\subsection{Physical calculation}
	\vspace*{-2mm}
	We create the \ac{RSS} map for the cell of interest using numerical methods. A common tool is \textit{Wireless Insite} \cite{remcom2022wireless}, which accurately models the electromagnetic environment. We assume the BS has a coarse estimate of the UE’s position, allowing us to extract approximate power levels and create the \ac{RSS} map for network input. In practice, \ac{RSS} maps can also be generated through digital twins that provide real-time virtual representations of the propagation environment \cite{ericsson2023digitaltwin}.
	
	\vspace*{-3mm}
	\subsection{Neural network design} 
\begin{figure}[!h]
		\centering
		\includegraphics[width=\linewidth]{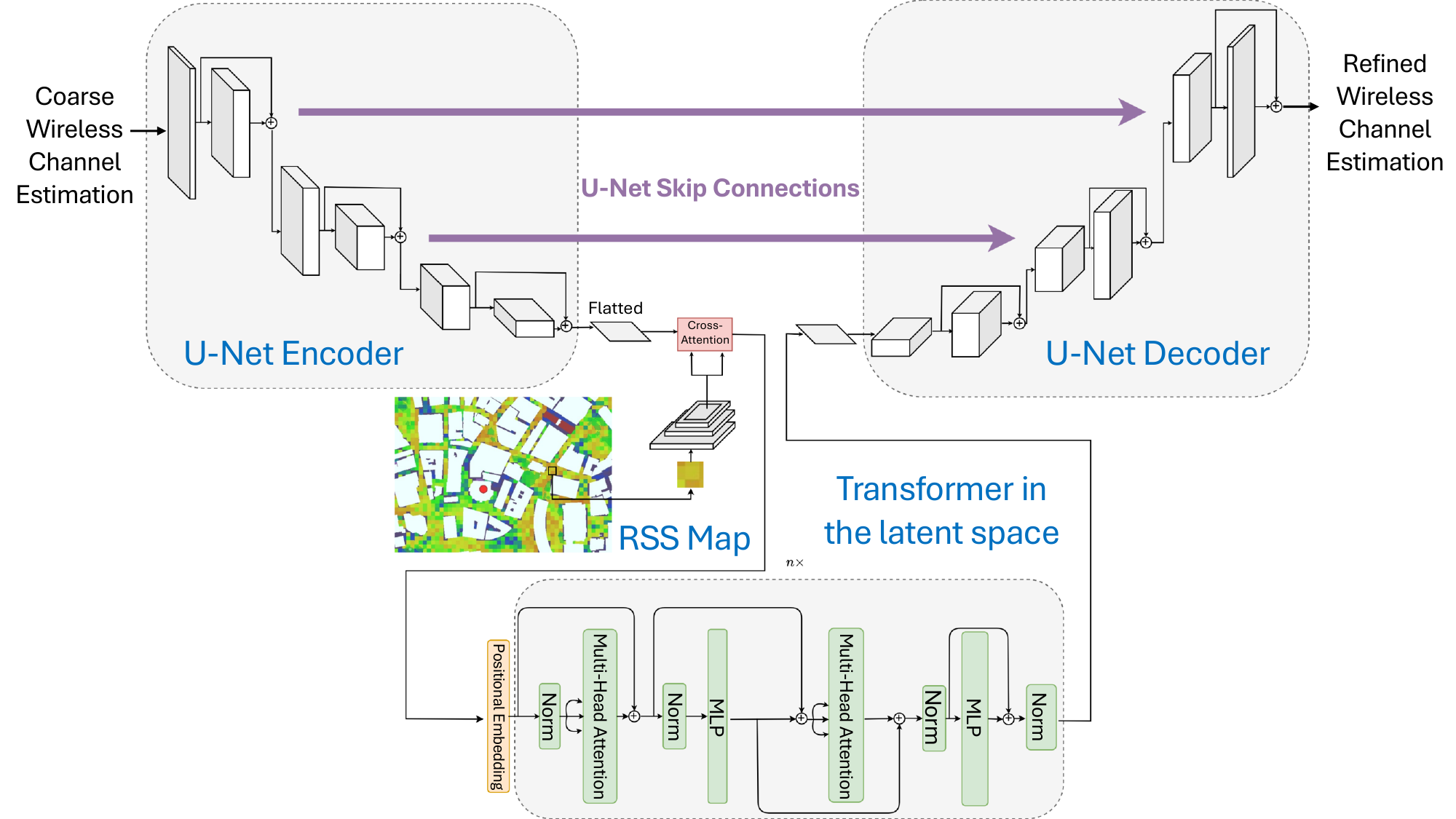}
		\caption{\ac{PINN} structure for channel estimation.}
		\label{fig:nn}
	\end{figure}

	\vspace*{-2mm}
	We design a physics-informed U-Net with transformer modules, as illustrated in Fig.~\ref{fig:nn}. The three-layer encoder–decoder uses ResNet blocks with skip connections to process the $2\times D$ channel input. In the latent space, features are fused with \ac{RSS} embeddings via cross-attention, enabling the network to integrate environmental propagation information with channel structure. The process starts with a projection to a common hidden dimension as
	\vspace*{-1mm}
	\begin{equation}
		\mathbf{X}_i = \mathbf{W}_i \mathbf{F}_i + \mathbf{b}_i\:\:; \:\: i \in \{\text{\small{RSS, Channel}}\}, 
	\end{equation}
	where $\mathbf{F}_i \in \mathbb{R}^{\text{Batch} \times D_i}$ is the extracted features from \ac{RSS} and channel, $\mathbf{X}_i \in \mathbb{R}^{\text{Batch} \times D_z}$ is the projected features, $\mathbf{W}_i  \in \mathbb{R}^{D_i \times D_z}$ is the projection matrices, and $\mathbf{b}_i\in \mathbb{R}^{ D_z \times 1}$ is the bias vector. Here, $D_z$ represents the dimension of the latent space and $D_i$ is the dimension of \ac{RSS} or channel features. The cross-attention mechanism employs multi-head attention. The process is formulated as $\text{softmax}\left(\frac{\mathbf{Q}\mathbf{K}^T}{\sqrt{D_z}}\right)\mathbf{V}$ where $\mathbf{Q} = \mathbf{X}_{\text{Channel}}$, $\mathbf{K} = \mathbf{X}_{RSS}$, and $\mathbf{V} = \mathbf{X}_{RSS}$. Setting $\mathbf{Q}$ as channel features, and $\mathbf{K}$, $\mathbf{V}$ as \ac{RSS} features, enables the channel estimation to query what environmental information is most relevant for refinement, where the attention weights determine which \ac{RSS} spatial patterns are most informative for each channel element. This cross-modal attention allows the network to selectively incorporate physics-based environmental knowledge into the channel estimation process rather than treating them as independent modalities.
	The fused features are refined by a transformer with self-attention blocks to capture long-range dependencies and physics-informed relations between \ac{RSS} patterns and channel characteristics. The decoder mirrors the encoder with transposed convolutions and skip connections to recover the channel dimensions while preserving details. This design combines U-Net’s spatial preservation with transformer modeling of channel–environment interactions. Implementation details are given in Section~\ref{imp}. Training uses a physics-informed loss that couples Maxwell-based constraints with reconstruction error 
	$    \mathcal{L}_{\text{total}} = \mathcal{L}_{\text{NMSE}} + \zeta \: \mathcal{L}_{\text{phy}}\: , $
	where $ \mathcal{L}_{\text{NMSE}}$ is the reconstruction loss for the channel estimation defined as
	\vspace*{-2mm}
	\begin{equation}
		\mathcal{L}_{\text{NMSE}} = \mathbb{E} \left( \frac{\|\mathbf{H - \hat{H}}\|_2^2}{\|\mathbf{H}\|_2^2}\right)
	\end{equation}
	in which $\mathbf{H} \in \mathbb{C}^{D \times N_{\text{r}} \times N_{\text{t}}}$ and $\mathbf{\hat{H}} \in \mathbb{C}^{D \times N_{\text{r}} \times N_{\text{t}}}$ are the 3-dimensional complex tensors for accurate and estimated channels respectively. Moreover, $\mathcal{L}_{\text{phy}}$ defined as
	%\begin{align}
	%\mathcal{L}_{\text{physical}}
	%&= \mathbb{E}\Bigg(
	%    \mathbb{E}\!\left[
	%        \frac{\lambda^{2}}{8\pi\eta_{0}}
	%        \left|\sum_{i=1}^{P} E_i(\mathbf r)\right|^{2}
	%    \right] \nonumber \\[0.5ex]
	%&\qquad -\, P_T
	%    \sum_{d=1}^{D}
	%    \mathbb{E}\!\left[
	%        \operatorname{tr}\!\left(\|\mathbf H_d\|_2^2\right)
	%    \right]
	%\Bigg)^{2}.
	\begin{equation}
		\mathcal{L}_{\text{phy}} \hspace*{-1mm} = \hspace*{-1mm}
		\mathbb{E}\Bigg(  \hspace*{-1mm}  \mathbb{E}\!\left[
		\frac{\lambda^{2}}{8\pi\eta_{0}}
		\left|\sum_{i=1}^{P} E_i(\mathbf r)\right|^{2}
		\right]\hspace*{-1mm} - P_T
		\sum_{d=1}^{D}
		\mathbb{E}\!\left[
		\!\left(\|\mathbf {\hat{H}_d}\|_F^2\right)
		\right]\hspace*{-1mm}
		\Bigg)^{2}\hspace*{-1mm}.
		\label{eq:Lphysical}
	\end{equation}	
	and $\zeta$ weights the physical consistency term to prevent physics violations while maintaining reconstruction accuracy.
	\vspace*{-2mm}
	\begin{remark}
		In the pilot-limited regime, the benefit of including the physics-informed term
		$\mathcal{L}_{\text{phy}}$ can be understood through a simple bias--variance
		argument \cite{bishop2006pattern}. Without constraints, a data-driven estimator must search over a
		high-dimensional hypothesis space, leading to variance scaling as
		$\operatorname{Var}[f_\theta] \propto d_{\text{eff}}/M$, where $f_\theta$ is the neural estimator with parameters $\theta$, $d_{\text{eff}}$ is the effective model dimension (less than the model dimension $d$) and $M$ the number of pilots. By enforcing approximate physical
		constraints $\mathcal{P}(f_\theta(\mathbf{y})) \approx 0$, with $\by$ the observations (pilots and RSS), the search space is
		restricted to a lower-dimensional manifold, yielding
		\vspace*{-2mm}
		\begin{equation}
			\operatorname{Var}[f_\theta] \propto d_{\text{eff}}/M, 
			\qquad d_{\text{eff}} \ll d,
		\end{equation}
		thus reducing estimation variance with only a minor bias if the physical model
		is imperfect. From an optimization viewpoint, the additional quadratic penalty
		$\lambda\mathcal{L}_{\text{phy}}$ improves the conditioning of the loss
		landscape. Writing a gradient step as
		\vspace*{-2mm}
		\begin{equation}
			\theta_{t+1} = \theta_t - \eta\bigl(\nabla \mathcal{L}_{\text{data}} +
			\lambda \nabla \mathcal{L}_{\text{phy}} + \xi_t\bigr),
		\end{equation}
		where $\xi_t$ denotes stochastic noise, the Hessian gains an extra
		regularizing term, which lowers the effective condition number and accelerates
		convergence. Hence, the physics-informed loss improves both generalization and
		training stability.
	\end{remark}
	\begin{table*}[htbp]
		\centering
		\caption{Physics-Informed U-Net architecture and parameters. $\downarrow$ and $\uparrow$ represent the downsampling and upsampling layers.}
		\label{tab:architecture}
		\begin{adjustbox}{max width=\textwidth}
			\begin{tabular}{ccc|ccc|ccc}
				\hline
				\multicolumn{3}{c|}{\textbf{Encoder}} & \multicolumn{3}{c|}{\textbf{Latent}} & \multicolumn{3}{c}{\textbf{Decoder}} \\
				\hline
				\textbf{\#} & \textbf{Type} & \textbf{Output Size} &
				\textbf{\#} & \textbf{Type} & \textbf{Output Size} &
				\textbf{\#} & \textbf{Type} & \textbf{Output Size} \\
				\hline
				Input                   & Channel         & $32\times4\times576$ &
				1                       & RSS Encoder    & $256$ &
				Output                  & Channel        & $32\times4\times576$ \\
				1\,($\downarrow$)       & ResUNetBlock    & $64\times2\times288$ &
				2                       & Cross-Attention& $18\,432$ &
				1\,($\uparrow$)         & ResUNetBlock   & $(128+128)\times1\times144$ \\
				2\,($\downarrow$)       & ResUNetBlock    & $128\times1\times144$ &
				3                       & Transformer    & $72\times256$ &
				2\,($\uparrow$)         & ResUNetBlock   & $(64+64)\times2\times288$ \\
				3\,($\downarrow$)       & ResUNetBlock    & $256\times1\times72$ &
				&                & &
				3\,($\uparrow$)         & ResUNetBlock   & $32\times4\times576$ \\
				\hline
			\end{tabular}
		\end{adjustbox}
	\end{table*}
	
	\vspace*{-5mm}
	\section{Numerical Experiments} \label{exps}
	\vspace*{-3mm}
	The codes  and dataset used to obtain the experimental results presented in this section are provided in \cite{codespinnwce}.
	%at \href{https://anonymous.4open.science/r/Physics-Informed-Neural-Networks-for-Wireless-Channel-Estimation-with-Limited-Pilot-Signals-2DB4/README.md}{GitHub}.
	\vspace*{-4mm}
	\subsection{Dataset}
	\vspace*{-2mm}
	We generated around 10,000  urban channels in Boston using Wireless InSite \cite{remcom2022wireless}. The BS was placed on a tall building, and vehicular UEs followed multiple trajectories with different velocities. Each multipath component included amplitude, delay, and angles, and a raised-cosine filter was considered to simulate filtering effects. The RSS maps were also produced to inject physical knowledge into the network. To test adaptability, we constructed datasets at both 8 and 15 GHz, with corresponding bandwidths of 200 and 400 MHz, respectively.
	\vspace*{-4mm}
	\subsection{Network Implementation} \label{imp}
	\vspace*{-2mm}
	Figure~\ref{fig:nn} shows the encoder, latent domain, and decoder of our network, with layer sizes given in Table~\ref{tab:architecture}. We use ResUNet blocks that merge U-Net skip connections with ResNet residuals \cite{ronneberger2015u, he2016deep}, improving feature propagation and gradient stability for channel estimation.
	We process \ac{RSS} inputs with a compact CNN encoder that extracts spatial patterns into a 256‐dimensional embedding for channel estimation. It uses stacked $3\times3$ convolutions with ReLU, $2\times2$ max‐pooling, and a final adaptive pooling layer. 
	\vspace*{-5mm}
	\subsection{Numerical Results}
\begin{figure}[!h]
		\centering
		\begin{subfigure}[b]{0.495\linewidth}
			\centering
			\includegraphics[width=\linewidth]{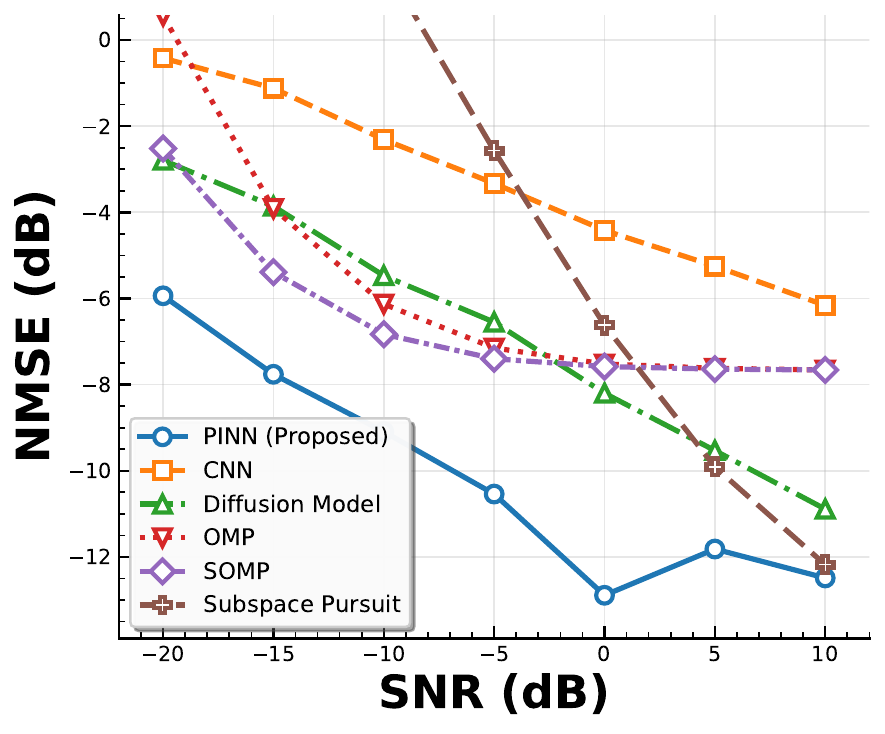}
			\caption{NMSE vs.\ SNR with $N_p = 4$ ($N_p=64$ for classical methods).}
			\label{fig:per2}
		\end{subfigure}
		\hfill
		\begin{subfigure}[b]{0.495\linewidth}
			\centering
			\includegraphics[width=\linewidth]{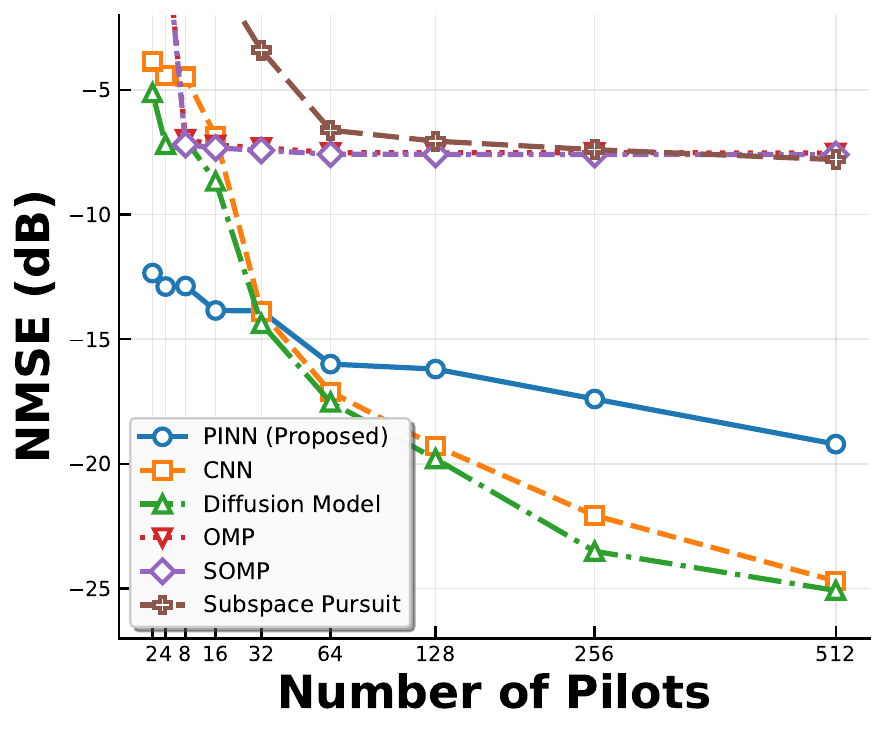}
			\caption{NMSE vs.\ number of pilot signals at $\mathrm{SNR}=0$ dB.}
			\label{fig:per3}
		\end{subfigure}
		\vspace*{-2mm}
		\caption{NMSE comparison across varying SNRs and pilot counts.}
		\label{fig:combined_nmse}
	\end{figure}

	\vspace*{-3mm}
	Using the hyperparameters in Table~\ref{tab:training_params}, we trained with an 80/10/10 train/val/test split. All samples and RSS values were normalized by a fixed constant for consistent NMSE loss scaling. Unless noted otherwise, experiments use $N_t=24\times24$, $N_r=2\times2$, and $D=16$. We also modeled the channel at an upper-mid band frequency of \(f_c = 15\) GHz \cite{kang2024cellular}. We set the number of subcarriers $N=1024$ and change the subcarrier spacing, and accordingly, the number of pilot signals. We compared the proposed PINN with classical OMP methods\cite{dai2009subspace, lee2016channel, tropp2007signal}, a CNN-based estimator \cite{sattari2024full}, and a diffusion model \cite{zhou2025generative}. We focus on performance testing at low-SNR conditions \cite{jin2025near} in Figure~\ref{fig:per2}. We use $N_p=4$ pilots  for PINN and $N_p=64$ for classical methods. As shown in Figure~\ref{fig:per2}, PINN outperforms all baselines,  achieving $\sim$5 dB NMSE gain at $\mathrm{SNR}=0$, demonstrating robustness in scarce-pilot regimes and the benefit of leveraging environmental information. Moreover, Table~\ref{tab:pinn-comparison} highlights the key complexity analysis for the proposed method and \cite{zhou2025generative}. This shows that \ac{PINN} requires significantly fewer FLOPs and achieves much lower inference latency. The diffusion architecture, however, maintains a low parameter, making it more memory‑efficient.
	Figure~\ref{fig:per3} shows NMSE vs.\ pilot count at $\mathrm{SNR}=0$. CNN and diffusion methods perform well with many pilots, but PINN clearly outperforms in pilot-limited regimes. 
	%This highlights the value of physics-informed inductive bias, which compensates for sparse pilot information and ensures robust estimation.
	%% TEST 2
	\vspace*{-4mm}
	\begin{table}[h!]
		\centering
		\caption{Training hyperparameters}
		\label{tab:training_params}
		\begin{tabular}{lcccc}
			\toprule
			Batch & Epochs & LR & Optimizer & $\zeta$ \\
			\midrule
			32 & 500 & $10^{-3}$ & Adam & 0.01 \\
			\bottomrule
		\end{tabular}
	\end{table}
	\vspace*{-4mm}
	\begin{table}[h!]
		\centering
		\caption{Complexity, latency, and parameters comparison}
		\label{tab:pinn-comparison}
		\begin{tabular}{lccc}
			\toprule
			& \multicolumn{1}{c}{\textbf{FLOPs}} 
			& \multicolumn{1}{c}{\textbf{Latency [ms]}} 
			& \multicolumn{1}{c}{\textbf{Parameters}} \\
			\cmidrule(lr){2-2} \cmidrule(lr){3-3} \cmidrule(lr){4-4}
			$(N_t,N_r,D)$ 
			& \((576,4,16)\) 
			& \((576,4,16)\) 
			& \((576,4,16)\) \\
			\midrule
			PINN & 70.85G & 11.12 & $3.5 \times 10^8$  \\
			DM & 130.15G & 50.30 & $5.5 \times 10^4$\\
			\bottomrule
		\end{tabular}
	\end{table}
	
\begin{figure}[!h]
		\centering
		\begin{subfigure}[b]{0.49\linewidth}
			\centering
			\includegraphics[width=\linewidth]{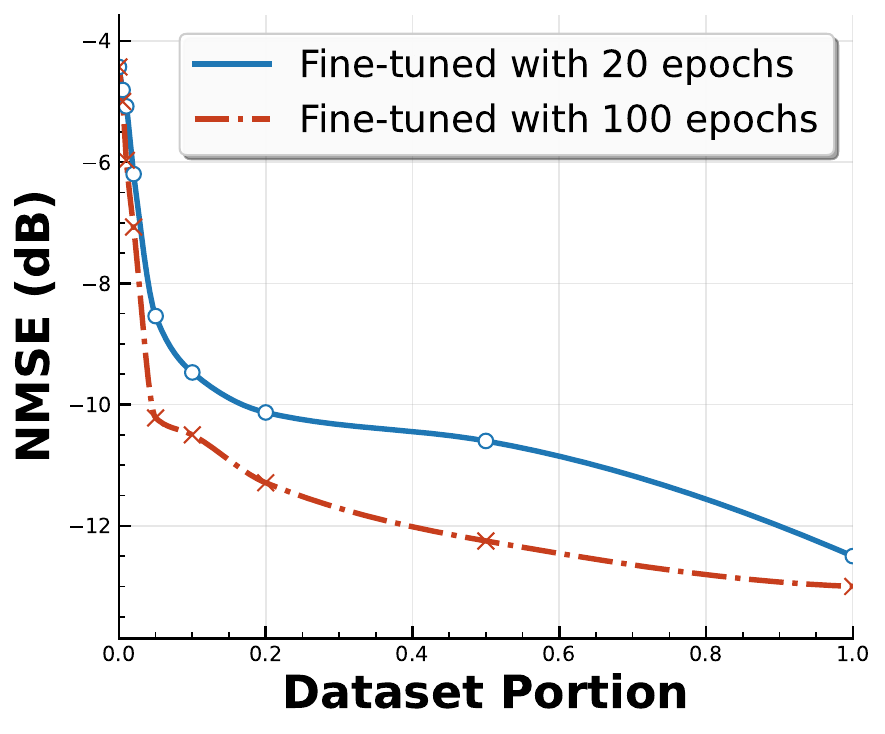} 
			\caption{From 15 GHz to 8 GHz in Boston environment}
			\label{fig:8GHZ}
		\end{subfigure}
		\hfill
		\begin{subfigure}[b]{0.49\linewidth}
			\centering
			\includegraphics[width=\linewidth]{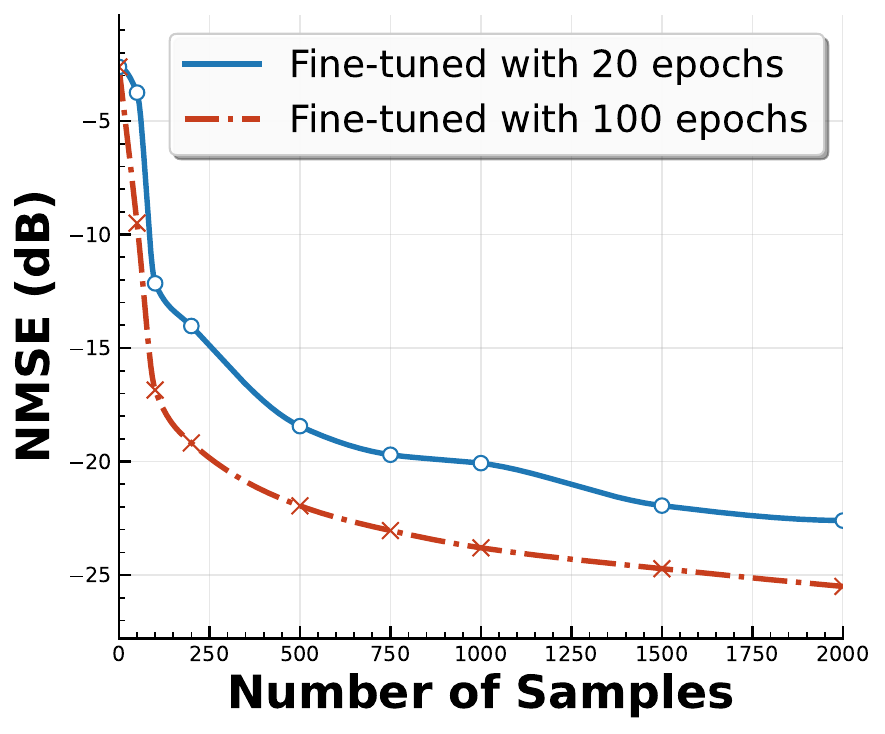}
			\caption{From Boston to urban canyon environment}
			\label{fig:canyon1}
		\end{subfigure}
		\caption{Transfer learning performance: NMSE vs. dataset for 20 and 100 epoch fine-tuning using $N_p=4$ and $\mathrm{SNR}=0$.}
		\label{fig:combinedse}
	\end{figure}

	\vspace*{-3mm}
	To evaluate the generalization ability, we tested transfer learning from 15 to 8 GHz (with 400 to 200 MHz bandwidth) using 4 pilots at $\mathrm{SNR}=0$ dB. As shown in Figure~\ref{fig:combinedse}, even 10\% of the 8 GHz data yields NMSE $\approx-5$ dB, while full training reaches $\approx-13$ dB. We also transferred from the Boston map to an urban canyon, where the model rapidly adapts: with only 100 samples it achieves NMSE $\approx-9$ dB (100 epochs), improving to below $-25$ dB with 2000 samples. These results confirm that the proposed PINN captures environment-agnostic propagation principles, enabling efficient adaptation across bands and scenarios with limited fine-tuning.
	
	\vspace*{-9mm}
	\section{Conclusion} \label{con}
	\vspace*{-3mm}
	This paper introduced a \ac{PINN} for accurate wireless channel estimation under pilot-constrained scenarios. By integrating initial simple least-squares estimates with \ac{RSS} maps derived from Maxwell-based ray tracing, the proposed architecture leverages environmental knowledge through a U-Net backbone enhanced with transformer and cross-attention modules. Our results on realistic 15 GHz urban ray-traced data show that \ac{PINN} significantly outperforms conventional and learning-based baselines, especially in low-pilot and low-SNR regimes, achieving up to 15 dB \ac{NMSE} gain over initial \ac{LS} estimates and over 4 dB gain compared to state-of-the-art models. 
	%At the end, we investigated the generalization capability of this method in different frequency bands and environments. 
	With low inference latency and strong interpretability, \ac{PINN} offers a scalable and physically grounded solution for real-time deployment in next-generation \ac{MIMO} systems.
	
	\newpage
	
	\section{Acknowledgments}
	This material is based upon work partially supported by funds from the industry affiliate program at the Center for Wireless Communications (CWC) at UC San Diego.
	
	\bibliographystyle{IEEEtran}
	\bibliography{refs}
	
\end{document}

%% file: macros.tex
\usepackage{amsmath,amssymb}
\newcommand{\br}{{\mathbf{r}}}

\newcommand{\by}{{\mathbf{y}}}

\makeatletter
\def\munderbar#1{\underline{\sbox\tw@{$#1$}\dp\tw@\z@\box\tw@}}
\makeatother

%% file: acronyms.tex
\usepackage{acro}
\DeclareAcronym{3GPP}{
	short=3GPP,
	long=3rd Generation Partnership Project
}
\DeclareAcronym{ADC}{
	short=ADC,
	long=analog-to-digital converter
}
\DeclareAcronym{ADP}{
	short=ADP,
	long=angular-delay profile
}
\DeclareAcronym{AMP}{
	short=AMP,
	long=approximate message passing
}
\DeclareAcronym{AoA}{
	short=AoA,
	long=angle-of-arrival
}
\DeclareAcronym{AoD}{
	short=AoD,
	long=angle-of-departure
}
\DeclareAcronym{APS}{
	short=APS,
	long=azimuth power spectrum
}
\DeclareAcronym{AWGN}{
	short=AWGN,
	long=additive white Gaussian noise
}
\DeclareAcronym{AV}{
	short=AV,
	long=autonomous vehicle
}
\DeclareAcronym{BI}{
	short=BI,
	long=Bayesian inference
}
\DeclareAcronym{BS}{
	short=BS,
	long=base station
}
\DeclareAcronym{BSM}{
	short=BSM,
	long=basic safety message
}
\DeclareAcronym{CDF}{
	short=CDF,
	long=cumulative distribution function
}
\DeclareAcronym{CIR}{
	short=CIR,
	long=channel impulse response
}
\DeclareAcronym{CNN}{
	short=CNN,
	long=convolutional neural network
}
\DeclareAcronym{ChanSTA}{
	short=ChanSTA,
	long=channel spatial and temporal attention
}

\DeclareAcronym{CP}{
	short=CP,
	long=cyclic-prefix
}

\DeclareAcronym{CS}{
	short=CS,
	long=compressed sensing
}
\DeclareAcronym{CSI}{
	short=CSI,
	long=channel state information
}

\DeclareAcronym{DFT}{
	short=DFT,
	long=discrete Fourier transform
}
\DeclareAcronym{IFFT}{
	short=IFFT,
	long=inverse fast Fourier transform
}
\DeclareAcronym{FFT}{
	short=FFT,
	long=fast Fourier transform
}
\DeclareAcronym{DL}{
	short=DL,
	long=deep learning
}

\DeclareAcronym{DNN}{
	short=DNN,
	long=deep neural network
}
\DeclareAcronym{DoA}{
	short=DoA,
	long=direction-of-arrival
}
\DeclareAcronym{DoD}{
	short=DoD,
	long=direction-of-departure
}
\DeclareAcronym{DSRC}{
	short=DSRC,
	long=dedicated short-range communication
}
\DeclareAcronym{EM}{
	short=EM,
	long=expectation maximization
}
\DeclareAcronym{FC}{
	short=FC,
	long=fully connected
}

\DeclareAcronym{FDD}{
	short=FDD,
	long=frequency division duplex
}
\DeclareAcronym{FMCW}{
	short=FMCW,
	long=frequency modulated continuous wave
}
\DeclareAcronym{FoV}{
	short=FoV,
	long=field-of-view
}
\DeclareAcronym{GNSS}{
	short=GNSS,
	long=global navigation satellite system
}
\DeclareAcronym{GPS}{
	short=GPS,
	long=global positioning system
}
\DeclareAcronym{IoT}{
	short=IoT,
	long=internet of things
}
\DeclareAcronym{IMU}{
	short=IMU,
	long=inertial measurement unit 
}
\DeclareAcronym{KL}{
	short=KL,
	long=Kullback–Leibler
}
\DeclareAcronym{KF}{
	short=KF,
	long=Kalman filter
}
\DeclareAcronym{LIDAR}{
	short=LIDAR,
	long=light detection and ranging
}
\DeclareAcronym{LOS}{
	short=LOS,
	long=line-of-sight
}
\DeclareAcronym{LPF}{
	short=LPF,
	long=low pass filter
}
\DeclareAcronym{LTE}{
	short=LTE,
	long=long term evolution
}
\DeclareAcronym{LS}{
	short=LS,
	long=least squares
}
\DeclareAcronym{LSTM}{
	short=LSTM,
	long=long short-term memory
}
\DeclareAcronym{mmWave}
{
	short = mmWave, 
	long = millimeter wave
}
\DeclareAcronym{MOMP}{
	short=MOMP,
	long=multidimensional orthogonal matching pursuit
}
\DeclareAcronym{MUSIC}{
	short=MUSIC,
	long=multiple signal classification
}
\DeclareAcronym{MIMO}{
	short=MIMO,
	long=multiple-input multiple-output
}
\DeclareAcronym{MHA}{
	short=MHA,
	long=multi-head attention
}
\DeclareAcronym{ML}{
	short=ML,
	long=machine learning
}
\DeclareAcronym{MLE}{
	short=MLE,
	long=maximum likelihood estimation
}
\DeclareAcronym{MLP}{
	short=MLP,
	long=multilayer perceptron
}
\DeclareAcronym{MRR}{
	short=MRR,
	long=medium range radar
}
\DeclareAcronym{MSE}{
	short=MSE,
	long=mean squared error
}
\DeclareAcronym{NMSE}{
	short=NMSE,
	long=normalized mean squared error
}
\DeclareAcronym{NLOS}{
	short=NLOS,
	long=non-line-of-sight
}

\DeclareAcronym{NLP}{
	short=NLP,
	long=natural language processing
}

\DeclareAcronym{NR}{
	short=NR,
	long=new radio
}
\DeclareAcronym{OFDM}{
	short=OFDM,
	long=orthogonal frequency-division multiplexing
}
\DeclareAcronym{OMP}{
	short=OMP,
	long=orthogonal matching pursuit
}
\DeclareAcronym{PDP}{
	short=PDP,
	long=power delay profiles
}
\DeclareAcronym{PO}{
	short=PO,
	long=phase offset
}
\DeclareAcronym{ppm}{
	short=ppm,
	long=parts-per-million
}
\DeclareAcronym{RF}{
	short=RF,
	long=radio frequency
 }
\DeclareAcronym{RMS}{
	short=RMS,
	long=root-mean-square
}
\DeclareAcronym{RPE}{
	short=RPE,
	long=relative precoding efficiency
}
\DeclareAcronym{RSU}{
	short=RSU,
	long=roadside unit
}
\DeclareAcronym{RTT}{
	short=RTT,
	long=round trip time
}
\DeclareAcronym{RX}{
	short=RX,
	long=receiver
}

\DeclareAcronym{RSRP}{
	short=RSRP,
	long=reference signal received power
 }

\DeclareAcronym{SNR}{
	short=SNR,
	long=signal-to-noise ratio
}
\DeclareAcronym{SVD}{
	short=SVD,
	long=singular value decomposition
}
\DeclareAcronym{SLAM}{
	short=SLAM,
	long=simultaneous localization and mapping
}
\DeclareAcronym{SBL}{
	short=SBL,
	long=sparse Bayesian learning
}
\DeclareAcronym{SOMP}{
	short=SOMP,
	long=simultaneous orthogonal matching pursuit 
}

\DeclareAcronym{TCN}{
	short=TCN,
	long=temporal convolutional network
}
\DeclareAcronym{ToF}{
	short=ToF,
	long=time of flight
}
\DeclareAcronym{TX}{
	short=TX,
	long=transmitter
}

\DeclareAcronym{TDoA}{
	short=TDoA,
	long=time-difference-of-arrival
}

\DeclareAcronym{ToA}{
	short=ToA,
	long=time-of-arrival
}
\DeclareAcronym{TO}{
	short=TO,
	long=timing offset
}

\DeclareAcronym{UL}{
	short=UL,
	long=uplink
}
\DeclareAcronym{ULA}{
	short=ULA,
	long=uniform linear array 
}
\DeclareAcronym{URA}{
	short=URA,
	long=uniform rectangular array 
}
\DeclareAcronym{V2I}{
	short=V2I,
	long=vehicle-to-infrastructure
}
\DeclareAcronym{V2V}{
	short=V2V,
	long=vehicle-to-vehicle
}
\DeclareAcronym{V2X}{
	short=V2X,
	long=vehicle-to-everything
}
\DeclareAcronym{VRU}{
	short=VRU,
	long=vulnerable road user
}
\DeclareAcronym{WLS}{
	short=WLS,
	long=weighted least squares
}
\DeclareAcronym{ViT}{
	short=ViT,
	long=vision transformer
}
\DeclareAcronym{MIM}{
	short=MIM,
	long=masked image modeling
}
\DeclareAcronym{MPC}{
	short=MPCs,
	long=multipath components
}
\DeclareAcronym{RSS}{
	short=RSS,
	long=received signal strength
}
\DeclareAcronym{PINN}{
	short=PINN,
	long=physics-informed neural network
}